\documentstyle[psfig]{mn}
 
\def\a{\AA \ }
\def\.{\'\i}

\def\etal{{\it et al. }}
\def\abigg{\lower 3pt\hbox{$\buildrel > \over \sim\;$}}
\def\aless{\lower 3pt\hbox{$\buildrel < \over \sim\;$}}

\newcommand{\Hii}{\hbox{H\,{\sc ii}}\ }
\newcommand{\Oii}{[\hbox{O\,{\sc ii}}]\ } 
\newcommand{\Nii}{[\hbox{N\,{\sc ii}}]\ }
\newcommand{\Sii}{[\hbox{S\,{\sc ii}}]\ }
\newcommand{\Oiii}{[\hbox{O\,{\sc iii}}]\ }
\newcommand{\Siii}{[\hbox{S\,{\sc iii}}]\ }
\newcommand{\Siv}{[\hbox{S\,{\sc iv}}]\ }

\title[I.The N2 calibrator]
 {New light on the search for low metallicity galaxies \\
  I. The N2 calibrator}

\author[G.~Denicol\'o \etal]
   {Glenda Denicol\'o$^1$, Roberto Terlevich$^1$\thanks{Visiting Professor, 
INAOE, Puebla, Mexico} 
and Elena Terlevich,$^2$\thanks{Visiting Fellow, IoA, Cambridge} 
  \\
$^1$Institute of Astronomy, Madingley Road, Cambridge, CB3 0HA, United 
Kingdom\\
$^2$Instituto Nacional de Astrof\'{\i}sica, \'Optica y Electr\'onica, 
Tonantzintla, Puebla, Mexico}

\pagerange{\pageref{firstpage}--\pageref{lastpage}}
\pubyear{2001}

\begin{document}

\maketitle

\label{firstpage}


\begin{abstract}
We present a simple metallicity estimator based on the 
logarithmic \Nii$\lambda$6584\AA /H$\alpha$ ratio, hereafter N2, which we envisage will
become very useful for ranking galaxies in a metallicity sequence from redshift survey 
quality data even for moderately low spectral resolution.

We have calibrated the
N2 estimator using a compilation of \Hii galaxies having
accurate oxygen abundances, plus photoionization models
covering a wide range of abundances. The comparison of models and observations
indicates that both primary and secondary nitrogen 
are important for the relevant range of metallicities.

The N2 estimator follows a linear relation with log(O/H) that holds 
for the whole abundance range covered by the sample,  from
about 1/50th to twice the Solar value (7.2 $<$ 12+log(O/H) $<$ 9.1). We suggest that the 
(\Sii$\lambda\lambda$6717,6731\AA /H$\alpha$) ratio (hereafter S2) can also be used 
as a rough metallicity indicator. Because of its large scatter the S2 
estimator will be useful only in systems 
with very low metallicity, where \Nii$\lambda$6584\AA \  is not detected or
in low resolution spectra where \Nii$\lambda$6584\AA \  is blended with
H$\alpha$.

\end{abstract}


\begin{keywords}
galaxies: abundances -- galaxies: stellar content -- galaxies: evolution 
\end{keywords}

\section{Introduction}

The key for accurate determination of oxygen abundance in gaseous ionized
nebulae is a precise measurement of the weak auroral forbidden 
emission 
line \Oiii $\lambda$4363\AA . In  narrow emission line starforming galaxies
the temperature sensitive \Oiii $\lambda$4363\AA\  line intensity correlates with
the overall abundance, being relatively strong in very low metallicity
systems (12+log(O/H) $<$ 7.8) and becoming undetectable even for moderately 
low metallicity galaxies (12+log(O/H) $>$ 8.3). 
As a result, for most of the starforming  galaxies 
\Oiii $\lambda$4363\AA\  is unmeasurably weak. 

For the large majority of starforming regions the oxygen abundance 
is therefore estimated using empirical methods based on the relative 
intensities 
of strong, easily observable, optical lines. Although abundances 
derived in this way are recognized to 
suffer considerable uncertainties, still they are believed 
to be able to roughly trace general
trends in galaxies. The most widely used empirical abundance calibrators 
are the R$_{23}$ (Pagel \etal\  1979) and, recently, the S$_{23(4)}$ 
(V\.lchez \& Esteban 1996; D\.az \& P\'erez-Montero 2000; Oey \& Shields 2000)
parameters.

The R$_{23}$ method was first proposed by Pagel \etal (1979) and subsequently 
developed and calibrated by many authors (Edmunds \& Pagel 1984, McCall, 
Rybski \& Shields 1985, Dopita \& Evans 1986, Torres-Peimbert, Peimbert \& 
Fierro 1989, McGaugh 1991). 
It is defined as the sum of the flux of  \Oii$\lambda$3727\AA \  and 
\Oiii$\lambda\lambda$4959,5007\AA \  lines
relative to H$\beta$ (R$_{23}$ = (\Oii $\lambda$3727\AA \  + 
\Oiii$\lambda\lambda$4959,5007\AA )/H$\beta$). The R$_{23}$ method can be
used to estimate abundances up to relatively high redshifts, but
there are a few problems associated with the use of this estimator. 
Firstly, it is bi-valued, i.e. 
a single value of R$_{23}$ can be due to two very different 
oxygen abundances. Secondly,  a very large 
fraction of the starforming regions lie on the ill-defined turning zone 
around 12+log(O/H) $\simeq$ 8.1 
where regions with the same R$_{23}$ value have oxygen abundances 
which differ by almost an order of magnitude. 
Thirdly, the  R$_{23}$ method requires spectrophotometric
data  and given the wavelength range covered, the reddening correction
of the lines involved becomes crucial.
Finally a characteristic 
which is readily apparent is the large scatter 
in the R$_{23}$ vs.~oxygen abundance calibration, larger 
than accounted for by observational errors (Kobulnicky, Kennicutt \& Pizagno 1999).

The S$_{23}$ parameter introduced by V\.lchez \& Esteban 
(1996) is defined as the sum of the flux of  
\Sii$\lambda\lambda$6717,6731\AA \  and \Siii$\lambda\lambda$9069,9532\AA \ 
relative to H$\beta$ 
(S$_{23}$ = (\Sii$\lambda\lambda$6717,6731\AA \  + 
\Siii$\lambda\lambda$9069,9532\AA )/H$\beta$).
Unlike the R$_{23}$ method, the relation between S$_{23}$ and 
oxygen abundance remains single valued up to a metallicity slightly 
higher than solar. D\.az \& P\'erez-Montero (2000, hereafter DP00) show an 
empirical calibration of the S$_{23}$ 
parameter with a somewhat reduced scatter as compared to that 
with R$_{23}$. This calibration
should be improved by the inclusion of high quality data at both the low and 
the high metallicity ends,
but otherwise looks very promising for metallicities up to solar. 
On the other hand, Oey \& Shields (2000) argued that the S$_{23}$ 
method is more sensitive to the ionization parameter than R$_{23}$, 
and propose to include the emission of \Siv to overcome the limitations 
of S$_{23}$, introducing the S$_{234}$ method. 
A problem with this estimator is
that good data is still scarce, and that the detection of the sulphur lines 
in the near-infrared is limited  
to galaxies with redshifts smaller than about 0.1. 

In all, the estimation of abundances of large number of galaxies, 
particularly in the slightly subsolar to oversolar range,  
is still a difficult problem. This metallicity regime contains most of the 
\Hii regions in early spiral galaxies and 
the inner regions of most late type galaxies, therefore  the 
description of the metallicity distribution in galaxies cannot be complete
without it.

Storchi-Bergmann \etal (1994) suggested the use of 
N2 = \Nii$\lambda$6584\AA /H$\alpha$ , as an abundance estimator.
Their calibration of the 
N2 vs.~O/H relation  was improved by Raimann \etal (2000) that 
proposed a new  calibration using the Terlevich \etal 
(1991, hereafter T91) sample. 

In the present  work we have used the best available abundance data 
for starforming galaxies in combination with photoionization models
to explore the 
usefulness of the N2 abundance estimator and to calibrate it in terms of 
metallicity. 

The data set is presented in Section 2. 
The new calibration of the N2 estimator
 with photoionization models is discussed in Sections 3 and 4.
The concluding remarks are given in Section 5.


\section{The Data}

We have compiled from the literature a sample 
with the best available  data on the lines of interest. 

The published data included in our sample was selected for having 
high signal-to-noise, spectral resolution better than $\sim$ 8 \a 
in order to separate \Nii$\lambda$6584\AA \  from H$\alpha$, 
easily accessible line strengths and errors.
In computing abundances and line ratios, we have adopted the 
c(H$\beta$) reddening factor published by the respective authors. 

The sample of low metallicity galaxies is based on a 
compilation by Kobulnicky \& Skillman (1996) and is not intended 
to be complete, but includes the majority of the best spectrophotometric data  available
for \Hii galaxies. 

The data for the metal-poor galaxy sample (i.e., with available 
\Oiii$\lambda$4363\AA \  intensity and 12+log(O/H) $<$ 8.4) are presented in 
Table 1, 
together with their corresponding observational errors propagated from the
emission line uncertainties quoted in the original references. 
We re-determined the electron temperatures and 
abundances applying the method and expressions
from Section 4 of Pagel \etal (1992) to the reddening corrected emission 
line fluxes. 
In Table 1, objects within references 1-17 had their metallicities 
computed using the \Oiii$\lambda$4363\AA \  line intensity.
Galaxies with more than one \Hii region may appear more than 
once in the table. The comparison of the tabulated values can give an idea 
of observational and systematic errors and possible spatial variations.
We have added to the sample from the literature, 
our measurements of six metal-poor galaxies discovered by us on the 
Anglo-Australian 4.0m telescope in  August 1996 and
August 1997 (Terlevich \etal 2001, 
hereafter Paper II). The total number of low metallicity objects with 12+log(O/H) computed using the \Oiii $\lambda$4363\AA \  line intensity is 108.

The sample of  metal-rich galaxies (128) includes data from 
DP00; Castellanos, D\.az \& Terlevich (2001, hereafter CDT01) and T91. For  
galaxies with weak or none \Oiii$\lambda$4363\AA \  emission
(i.e., 12+log(O/H) $>$ 8.4), 
we derived the metallicity using R$_{23}$ (the {\it upper} branch analytic 
expression by McGaugh, 
published in Kobulnicky \etal 1999) or S$_{23}$ (D\.az \& P\'erez-Montero 
2000) methods, depending on the availability of the
\Siii$\lambda\lambda$9069,9532\AA \  lines. 
The errors for the metallicity values derived by the R$_{23}$ parameter already account for the intrinsic 0.2 dex uncertainty of the method.
The 55 metal-rich galaxies included in Table 1 with
references 18-26 correspond to the 
compilation in DP00 and to CDT01 and
had their metallicities estimated by the R$_{23}$ or S$_{23}$ methods. The 
rest 
of the metal rich galaxies, as they are from T91, were not included in the 
table.

The whole range of oxygen abundances covers from $\sim$ 2\% solar 
(Solar is taken as 12+log(O/H)$_{\odot}$ = 8.91) for IZw18 to more than solar 
for some regions for which detailed modeling has been performed 
(see compilation in DP00).


\begin{table*}
\centering
\begin{minipage}{140mm}
\caption{Chemical abundances and \Nii/H$\alpha$ ratios.}
\begin{tabular}{@{}lccl|clccr@{}}
\hline
\multicolumn{1}{l}{Galaxy} &  
\multicolumn{1}{c}{12+log(O/H)}  &
\multicolumn{1}{c}{N2}  & 
\multicolumn{1}{l}{Ref.}  &  \multicolumn{1}{c}\vline &
\multicolumn{1}{l}{Galaxy} &  
\multicolumn{1}{c}{12+log(O/H)}  &
\multicolumn{1}{c}{N2}  & 
\multicolumn{1}{l}{Ref.} \\ 
\hline

I Zw 18nw   & 7.196$\pm$0.046 & -2.527$\pm$0.068 & 1 & \vline & 1437+370  & 7.971$\pm$0.064 & -1.678$\pm$0.022 & 4 \\
I Zw 18se   & 7.309$\pm$0.057 & -2.317$\pm$0.017 & 1 & \vline & Pox 139   & 7.983$\pm$0.083 & -1.669$\pm$0.123 & 12\\
SBS 0335-052W&7.285$\pm$0.051 & -2.100$\pm$0.119 & 2 & \vline & UM 462    & 7.984$\pm$0.063 & -1.676$\pm$0.059 & 10\\
UGCA 292-1  & 7.320$\pm$0.050 & -1.943$\pm$0.046 & 3 & \vline &N5253 A    & 7.991$\pm$0.069 & -1.106$\pm$0.010 & 10\\
UGCA 292-2  & 7.361$\pm$0.060 & -2.104$\pm$0.060 & 3 & \vline &N5253-6    & 8.025$\pm$0.066 & -1.053$\pm$0.020 & 13\\
0940+544 N  & 7.377$\pm$0.040 & -2.224$\pm$0.055 & 4 & \vline &N5253-5    & 8.094$\pm$0.067 & -1.123$\pm$0.021 & 13\\
HS 0822+3542& 7.395$\pm$0.045 & -2.260$\pm$0.040 & 5 & \vline &N5253-1    & 8.144$\pm$0.076 & -1.182$\pm$0.012 & 13\\
1159+545    & 7.473$\pm$0.045 & -2.067$\pm$0.040 & 4 & \vline &N5253-2    & 8.167$\pm$0.072 & -1.215$\pm$0.021 & 13\\
1415+437    & 7.522$\pm$0.047 & -1.922$\pm$0.014 & 6 & \vline &N5253-4    & 8.180$\pm$0.069 & -1.179$\pm$0.024 & 13\\
0832+699    & 7.587$\pm$0.052 & -1.982$\pm$0.015 & 4 & \vline &N5253 B    & 8.266$\pm$0.104 & -1.161$\pm$0.036 & 10\\
T1214-277   & 7.596$\pm$0.052 & -2.493$\pm$0.048 & 7 & \vline &T1304-386  & 8.001$\pm$0.067 & -1.508$\pm$0.031 & 10\\
UGC4483 S   & 7.601$\pm$0.055 & -1.982$\pm$0.019 & 8 & \vline &UM 469     & 8.001$\pm$0.080 & -1.181$\pm$0.045 & 10\\
1211+540    & 7.687$\pm$0.054 & -2.142$\pm$0.022 & 4 & \vline &{\bf 37-27}& 8.012$\pm$0.064 & -1.630$\pm$0.025 & 9 \\
1249+493    & 7.721$\pm$0.053 & -1.926$\pm$0.054 & 6 & \vline &1533+469   & 8.012$\pm$0.062 & -1.261$\pm$0.011 & 6 \\
{\bf 537-69}& 7.735$\pm$0.101 & -1.628$\pm$0.046 & 9 & \vline &Mrk 600    & 8.019$\pm$0.068 & -1.873$\pm$0.122 & 7 \\
T1304-353   & 7.745$\pm$0.055 & -2.263$\pm$0.116 & 10& \vline &1135+581   & 8.019$\pm$0.069 & -1.625$\pm$0.006 & 4 \\
{\bf 70-05b}& 7.751$\pm$0.087 & -1.719$\pm$0.086 & 9 & \vline &Pox 108    & 8.026$\pm$0.089 & -1.669$\pm$0.123 & 12\\
UM 461      & 7.795$\pm$0.057 & -2.251$\pm$0.022 & 7 & \vline &Pox 4 NW   & 8.031$\pm$0.082 & -1.675$\pm$0.123 & 12\\
C1543+091   & 7.796$\pm$0.059 & -1.939$\pm$0.109 & 10& \vline & 0946+558  & 8.034$\pm$0.067 & -1.645$\pm$0.014 & 4 \\
1331+493 N  & 7.819$\pm$0.057 & -1.940$\pm$0.027 & 4 & \vline & 0948+532  & 8.039$\pm$0.068 & -1.617$\pm$0.020 & 4 \\
1331+493 S  & 7.911$\pm$0.062 & -1.441$\pm$0.031 & 6 & \vline & T1345-420 & 8.049$\pm$0.067 & -1.716$\pm$0.048 & 10\\
1152+579    & 7.850$\pm$0.057 & -1.844$\pm$0.022 & 4 & \vline & II Zw 70  & 8.067$\pm$0.090 & -1.346$\pm$0.029 & 11\\
Mrk 36      & 7.865$\pm$0.099 & -1.827$\pm$0.087 & 11& \vline &T1334-326  & 8.088$\pm$0.072 & -1.928$\pm$0.106 & 10\\
{\bf 25-10 }& 7.865$\pm$0.076 & -1.140$\pm$0.154 & 9 & \vline & II Zw 40  & 8.104$\pm$0.077 & -1.708$\pm$0.024 & 10\\
Pox 120     & 7.868$\pm$0.082 & -1.745$\pm$0.123 & 12& \vline & II Zw 40  & 8.183$\pm$0.091 & -1.778$\pm$0.051 & 11\\
Mrk 36      & 7.872$\pm$0.064 & -1.901$\pm$0.100 & 10& \vline & N6822-HuX & 8.104$\pm$0.086 & -1.740$\pm$0.100 & 14\\
{\bf 46-17 }& 7.891$\pm$0.061 & -1.968$\pm$0.028 & 9 & \vline &T1004-296 SE& 8.117$\pm$0.072 & -1.290$\pm$0.020 & 10\\
C1148-203   & 7.896$\pm$0.063 & -1.775$\pm$0.029 & 10& \vline &T1004-296 NW& 8.204$\pm$0.083 & -1.386$\pm$0.016 & 10\\
Pox 105     & 7.904$\pm$0.081 & -1.669$\pm$0.123 & 12& \vline & Tol 35    & 8.121$\pm$0.087 & -1.547$\pm$0.123 & 12\\
{\bf 297-24}& 7.939$\pm$0.081 & -1.325$\pm$0.083 & 9 & \vline & T0633-415 & 8.144$\pm$0.072 & -1.373$\pm$0.150 & 10\\
C0840+120   & 7.940$\pm$0.063 & -1.615$\pm$0.056 & 10& \vline & T0633-415 & 8.144$\pm$0.071 & -1.391$\pm$0.008 & 7 \\
Fairall 30  & 7.949$\pm$0.064 & -1.634$\pm$0.022 & 10& \vline & Fairall 2 & 8.169$\pm$0.074 & -1.155$\pm$0.066 & 10\\
Tol 2       & 7.965$\pm$0.080 & -1.351$\pm$0.123 & 12& \vline & T1324-276 & 8.170$\pm$0.077 & -1.895$\pm$0.024 & 10\\

\hline

\end{tabular}

{\it References to the table.}

(1)Skillman \& Kennicutt 1993; (2)Lipovetsky \etal 1999; (3)van Zee 2000; (4)Izotov, Thuan \& Lipovetsky 1994;
(5)Kniazev \etal 2000; (6)Thuan, Izotov \& Lipovetsky 1995; (7)Pagel \etal 1992; (8)Skillman \etal 1994;
(9)Terlevich \etal 2001; (10)Campbell, Terlevich \& Melnick 1986; (11)Garnett 1990;
 (12)Kunth \& Joubert 1985; (13)Walsh \& Roy 1989; (14)Pagel, Edmunds \& Smith 1980. 

\end{minipage}
\end{table*}
\setcounter{table}{0}
\begin{table*}
\centering
\begin{minipage}{140mm}
\caption{\it continued}
\begin{tabular}{@{}lcclclccl@{}}
\hline
\multicolumn{1}{l}{Galaxy} &  
\multicolumn{1}{c}{12+log(O/H)}  &
\multicolumn{1}{c}{N2}  & 
\multicolumn{1}{l}{Ref.} & \multicolumn{1}{c}{\vline} &
\multicolumn{1}{l}{Galaxy} &  
\multicolumn{1}{c}{12+log(O/H)}  &
\multicolumn{1}{c}{N2}  & 
\multicolumn{1}{l}{Ref.} \\ 
\hline

T1457-262 A & 8.170$\pm$0.074 & -1.362$\pm$0.065 & 10& \vline & NGC3310 E   & 8.482$\pm$0.051 & -0.687$\pm$0.005 & 21 \\
T1457-262 B & 8.230$\pm$0.081 & -1.666$\pm$0.058 & 10& \vline & NGC3310 L   & 8.551$\pm$0.077 & -0.563$\pm$0.008 & 21 \\
T1008-286   & 8.174$\pm$0.073 & -1.471$\pm$0.078 & 10& \vline & NGC3310 M   & 8.483$\pm$0.374 & -0.824$\pm$0.013 & 21 \\
UCM1612+1308& 8.183$\pm$0.070 & -1.593$\pm$0.012 & 15& \vline & NGC7714 A   & 8.473$\pm$0.051 & -0.509$\pm$0.011 & 22 \\
C1409+120   & 8.186$\pm$0.075 & -1.301$\pm$0.114 & 10& \vline & NGC7714 N110& 8.553$\pm$0.179 & -0.305$\pm$0.044 & 22 \\
N4212 C20   & 8.192$\pm$0.090 & -1.093$\pm$0.027 & 16& \vline & NGC7714 B   & 8.280$\pm$0.051 & -0.873$\pm$0.011 & 22 \\
N4214 A6    & 8.221$\pm$0.077 & -1.058$\pm$0.027 & 16& \vline & NGC7714 C   & 8.265$\pm$0.167 & -0.859$\pm$0.070 & 22 \\
N4214 C6    & 8.374$\pm$0.087 & -1.197$\pm$0.027 & 16& \vline & NGC7714 N216& 8.607$\pm$0.199 & -0.322$\pm$0.036 & 22 \\
 T0440-381  & 8.212$\pm$0.074 & -1.447$\pm$0.054 & 10& \vline & M101 NGC5471& 8.182$\pm$0.119 & -1.640$\pm$0.157 & 23 \\
 Mrk 5      & 8.214$\pm$0.090 & -1.309$\pm$0.053 & 11& \vline & M101 NGC5471& 8.012$\pm$0.123 & -1.468$\pm$0.024 & 19 \\
 LMC II2    & 8.225$\pm$0.084 & -1.612$\pm$0.061 & 17& \vline & M101 NGC5471A&7.986$\pm$0.142 & -1.553$\pm$0.054 & 24 \\
 Mrk 67     & 8.226$\pm$0.094 & -1.721$\pm$0.068 & 11& \vline & M101 NGC5471B&8.295$\pm$0.146 & -1.095$\pm$0.019 & 24 \\
T1116-325   & 8.336$\pm$0.096 & -1.537$\pm$0.080 & 10& \vline & M101 NGC5471C&8.187$\pm$0.179 & -0.994$\pm$0.030 & 24 \\
M33 CC93    & 8.497$\pm$0.164 & -0.540$\pm$0.021 & 18& \vline & M101 NGC5471D&8.200$\pm$0.210 & -1.553$\pm$0.054 & 24 \\
M33 IC142   & 8.583$\pm$0.142 & -0.694$\pm$0.031 & 18& \vline & M101 NGC5471E&8.158$\pm$0.205 & -1.757$\pm$0.087 & 24 \\
M33 NGC595  & 8.407$\pm$0.049 & -0.816$\pm$0.012 & 18& \vline & M101 NGC5455& 8.353$\pm$0.235 & -0.925$\pm$0.026 & 24 \\
M33 NGC595  & 8.461$\pm$0.188 & -0.735$\pm$0.029 & 19& \vline & M101 NGC5455& 8.480$\pm$0.188 & -0.771$\pm$0.153 & 23 \\
M33 MA2     & 8.401$\pm$0.073 & -1.031$\pm$0.033 & 18& \vline & M101 NGC5461& 8.499$\pm$0.339 & -0.854$\pm$0.022 & 24 \\
M33 NGC604  & 8.414$\pm$0.061 & -0.902$\pm$0.003 & 18& \vline & M51 CCM72   & 9.095$\pm$0.215 & -0.493$\pm$0.013 & 25 \\
M33 NGC604  & 8.260$\pm$0.131 & -0.940$\pm$0.020 & 19& \vline & M51 CCM24   & 9.040$\pm$0.222 & -0.464$\pm$0.063 & 25 \\
M33 NGC588  & 8.348$\pm$0.099 & -1.553$\pm$0.033 & 18& \vline & M51 CCM10   & 8.973$\pm$0.206 & -0.413$\pm$0.043 & 25 \\
M33 NGC588  & 8.395$\pm$0.244 & -1.346$\pm$0.023 & 19& \vline & NGC628 H13  & 8.194$\pm$0.023 & -0.761$\pm$0.008 & 26 \\
M33 IC131   & 8.357$\pm$0.263 & -0.959$\pm$0.026 & 19& \vline & NGC628 H3   & 8.234$\pm$0.024 & -0.784$\pm$0.024 & 26 \\
NGC2403 VS35& 8.430$\pm$0.086 & -0.706$\pm$0.019 & 20& \vline & NGC628 H4   & 8.310$\pm$0.026 & -0.679$\pm$0.040 & 26 \\
NGC2403 VS24& 8.423$\pm$0.129 & -0.828$\pm$0.019 & 20& \vline & NGC628 H5   & 8.340$\pm$0.028 & -0.678$\pm$0.040 & 26 \\
NGC2403 VS38& 8.365$\pm$0.150 & -0.860$\pm$0.019 & 20& \vline & NGC925 CDT1 & 8.520$\pm$0.038 & -0.693$\pm$0.029 & 26 \\
NGC2403 VS44& 8.332$\pm$0.125 & -0.876$\pm$0.019 & 20& \vline & NGC925 CDT2 & 8.719$\pm$0.051 & -0.602$\pm$0.037 & 26 \\
NGC2403 VS51& 8.353$\pm$0.196 & -1.012$\pm$0.020 & 20& \vline & NGC925 CDT3 & 8.505$\pm$0.041 & -0.639$\pm$0.035 & 26 \\
NGC2403 VS3 & 8.333$\pm$0.192 & -0.948$\pm$0.019 & 20& \vline & NGC925 CDT4 & 8.414$\pm$0.032 & -0.682$\pm$0.018 & 26 \\
NGC2403 VS49& 8.362$\pm$0.157 & -1.106$\pm$0.021 & 20& \vline & NGC1232 CDT1& 8.440$\pm$0.032 & -0.441$\pm$0.012 & 26 \\
NGC3310 Nuc & 8.831$\pm$0.059 & -0.298$\pm$0.007 & 21& \vline & NGC1232 CDT2& 8.597$\pm$0.041 & -0.602$\pm$0.024 & 26 \\
NGC3310 A   & 8.212$\pm$0.023 & -0.797$\pm$0.003 & 21& \vline & NGC1232 CDT3& 8.519$\pm$0.037 & -0.508$\pm$0.027 & 26 \\
NGC3310 B   & 8.441$\pm$0.032 & -0.638$\pm$0.007 & 21& \vline & NGC1232 CDT4& 8.516$\pm$0.037 & -0.578$\pm$0.023 & 26 \\
NGC3310 C   & 8.482$\pm$0.035 & -0.719$\pm$0.003 & 21& \vline & NGC1637 CDT1& 8.226$\pm$0.024 & -0.438$\pm$0.026 & 26 \\

\hline

\end{tabular}

{\it References to the table.}

(10)Campbell, Terlevich \& Melnick 1986; (11)Garnett 1990; (15)Rego \etal 1998; (16)Kobulnicky \& Skillman 1996; 
(17)Mathis, Chu \& Peterson 1985; (18)V\.lchez \etal 1988; (19)Garnett 1989; (20)Garnett \etal 1997; (21)Pastoriza \etal 1993; (22)Gonz\'alez-Delgado \etal 1994; (23)Shields \& Searle 1978; (24)Kennicutt \& Garnett 1996; (25)D\.az \etal 1991; (26)Castellanos, D\.az \& Terlevich 2001.    

\end{minipage}
\end{table*}

\section{Results}

The N2 parameter is defined as 

\begin{equation}
$N2 = log(\Nii$\lambda$6584/H$\alpha$).$ 
\end{equation}

\begin{figure*}
\vspace{12.5 cm}
\includegraphics{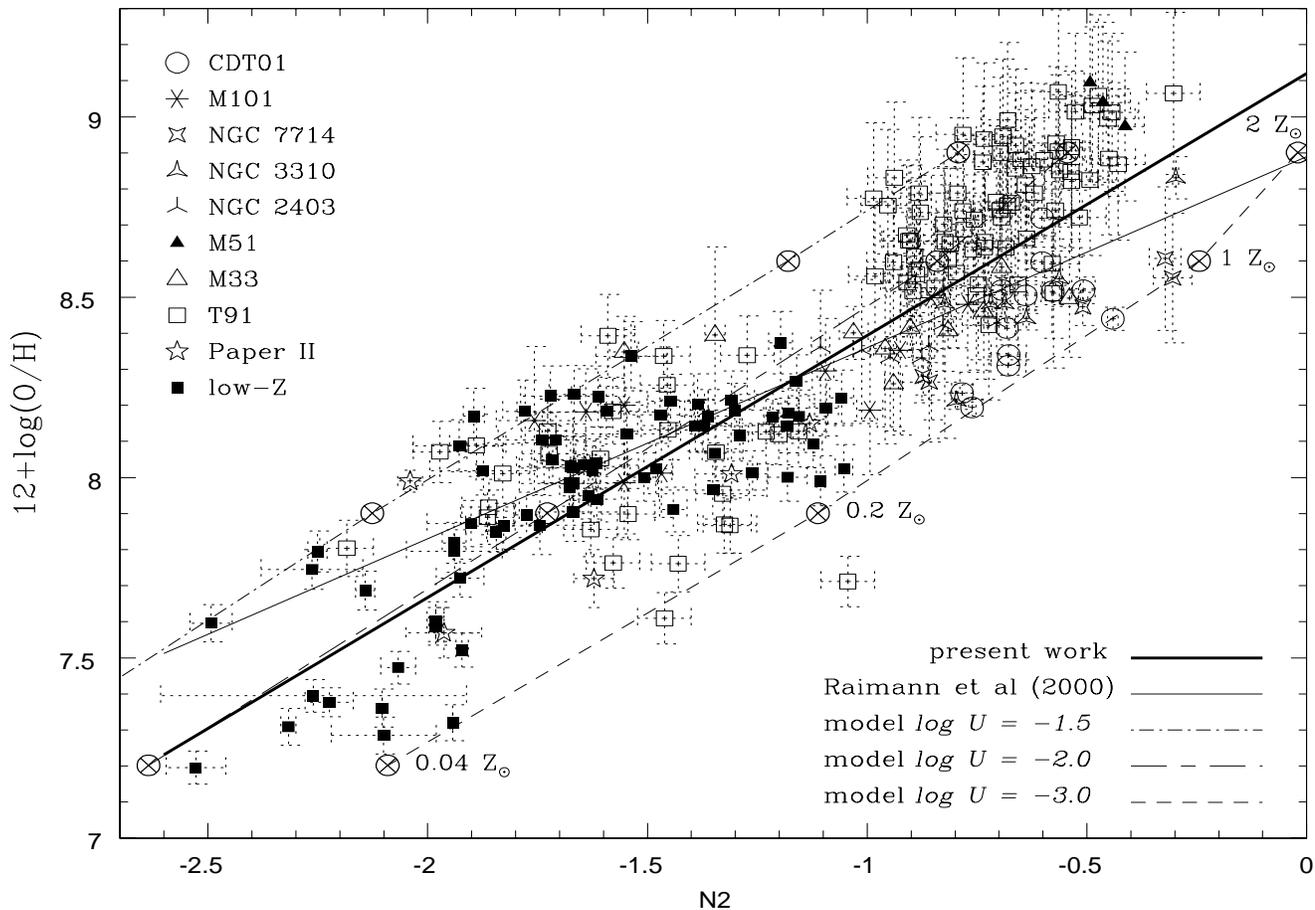}
\caption{Oxygen abundance vs.~the N2 calibrator (see text) for the whole 
dataset. The linear fit to the data, heavy solid line,
taking into account the dispersion in the
 distribution, gives 12+log(O/H) = 9.12+0.73$\cdot$N2, 
with a correlation coefficient of 0.85. The crossed-circle dots correspond to the photoionization model points for the metallicities as labelled by the side of the dots. The symbols refer to the galaxies in Castellanos, D\.az \& Terlevich 2001 (CDT01); six metal-poor galaxies from Terlevich \etal 2001 (Paper II), and a selected sample of low metallicity galaxies with references in Table 1. The only galaxies not listed in Table 1 come from the catalogue of \Hii galaxies of Terlevich \etal 1991 (T91).}
\end{figure*}

The relation between N2 and the oxygen abundance is shown in
Figure 1. It can be seen that these two parameters are well
correlated (linear correlation coefficient of 0.85) and that
a single slope is capable of describing the  
whole metallicity range, from the most metal-poor  
to the most metal-rich galaxies in the sample.  
The heavy solid line represents the linear fit to the 
N2 vs.~12+log(O/H) relation.  Least squares 
fits to the data simultaneously minimizing the errors in both axes, give

\begin{equation}
$12+log(O/H) = 9.12($\pm$ 0.05) + 0.73($\pm$ 0.10)$\cdot$N2 ,$ 
\end{equation}

The thin solid line in Figure 1 shows the linear fit from Raimann \etal (2000).
Given the differences among our samples and the smaller number of points
in Raimann \etal data, we conclude that both solutions are in basic
agreement. However both fits 
dramatically disagree with the calibration proposed by Storchi-Bergmann 
\etal (1994, section 4).

\subsection{Photoionization models}

We have computed a grid of photoionization 
models using CLOUDY (Ferland 1996) for abundances 
[O/H] = 0.04, 0.2, 1 
and 2. All the  elements apart from N are assumed to be primary
and are therefore  scaled as oxygen. For nitrogen we
used [N/O] = 0.08 + [O/H] that represents a realistic combination of 
early primary N plus later secondary N and reproduces the observed
behaviour of N/O with O/H (Henry, Edmunds \& K\"oppen, 2000
and references therein).
The square brackets in the expressions denote the value relative to that of Orion, i.e.,  N: 7.00 $\times$ 10$^{-5}$,  O: 4.00 $\times$ 10$^{-4}$.
For simplicity we have used single star models with 
T$_{eff} =$~45000~K. Changing to cluster models or including
a metallicity dependent T$_{eff} $ do not change the basic results.
The photoionization models are shown as dashed and dot-dashed
 lines in Figure 1. The lines join models with the same ionization
parameter ({\it U}). The whole range of the data is comprised between the models with 
{\it log (U)=-1.5} and {\it log (U)=-3.0}.
It can be seen that within the errors there is a good superposition
between models and observations.

Models with either pure primary or pure secondary nitrogen fail to 
reproduce the slope of the relation. The change in N2 per unit O/H change is too large 
when compared with the data, for the pure secondary N models, while it is too small
for the pure primary ones.

The abundance parameter N2 vs.~the ionization parameter sensitive ratio (for 
ionizing temperatures higher than about 35000 K) log (\Oii/\Oiii), 
is shown in Figure 2, together
with the photoionization model results. The lines represent again photoionization models
with the same ionization parameter and covering the whole range of metallicity.
As can be seen, the lines for constant ionization parameter are almost vertical
and at a given metallicity the range in N2 covered by changing the ionization 
parameter is much smaller than the observed range. 
This confirms that most of the observed trend of N2 with O/H is due to 
metallicity changes.

\begin{figure}
\vspace{8.0 cm}
\includegraphics{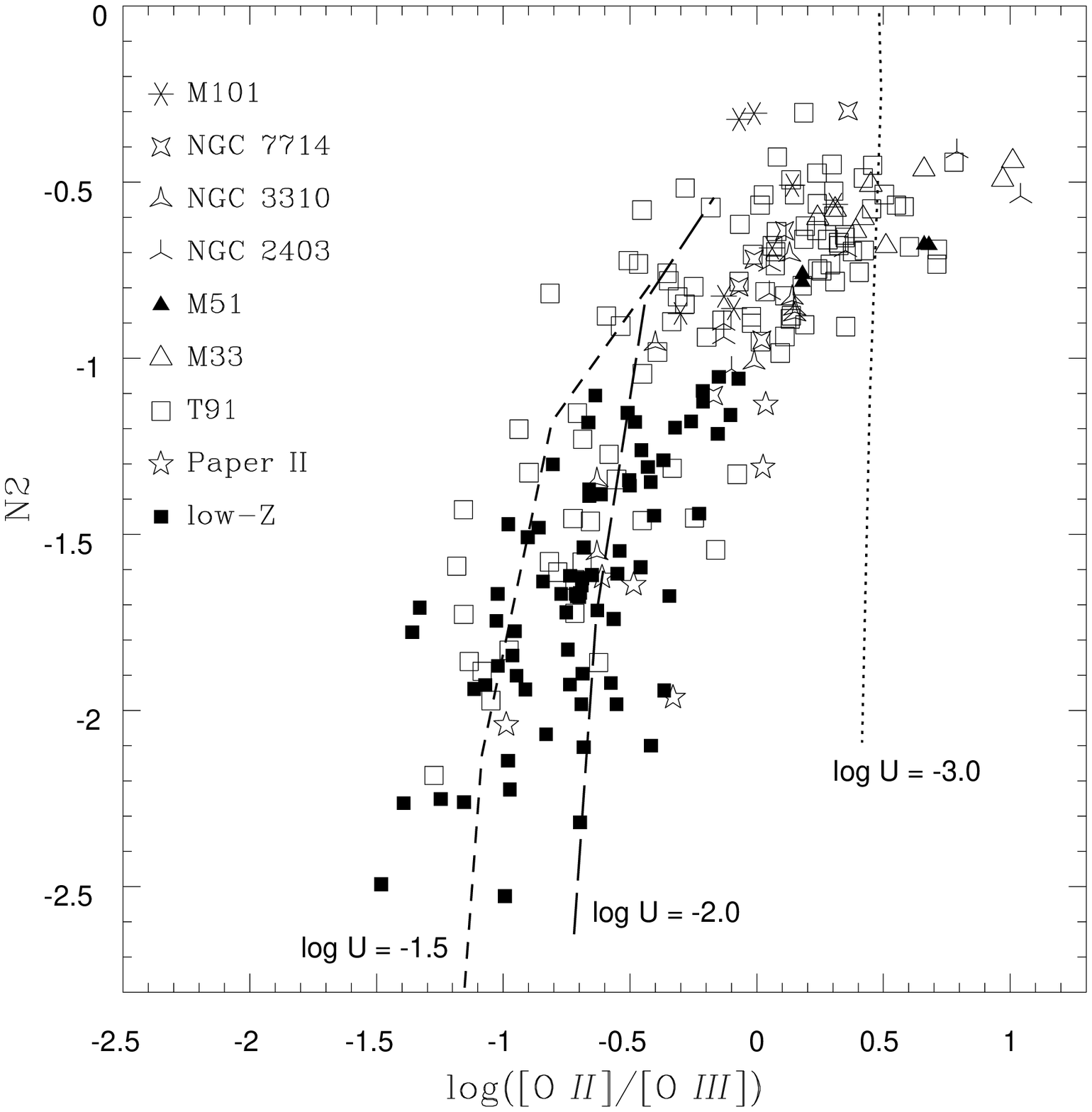}
\caption{The N2 parameter plotted against the ionization sensitive parameter
log (\Oii/\Oiii). The lines join models with the same ionization
parameter for abundances [O/H] = 0.04, 0.2, 1, 2 Z$_{\odot}$.}
\end{figure}

\section{Discussion}

Although relatively tight, the relation between N2 and abundance
has a scatter that might be larger than the observational errors.

The comparison of models and data in Figure 2 suggests that 
some of the scatter in the 
N2 vs.~12+log(O/H) relation could in principle depend on the 
degree of ionization of the nebula that in turn 
may depend on the age of the ionizing cluster.
To cover the whole data range 
in the N2 vs.~O/H relation, it was needed to vary the 
ionization parameter {\it U} from {\it log(U)}$=$-3 to -1.5, with a best fit 
corresponding to  {\it log(U)} $\cong$ -2. 

In an attempt to understand the dispersion in the O/H -- N2 relation, we have 
studied the connection between the scatter of the N2 parameter and 
an age sensitive parameter, the equivalent 
width of H$\beta$, EW(H$\beta$).  The ranking of  EW(H$\beta$) with age 
(Copetti, Pastoriza \& Dottori, 1986; Leitherer \& Heckman 1995)
goes in the sense that  EW(H$\beta$) in emission reaches values of several 
hundred \AA\ for an \Hii region photoionized by a zero-age coeval stellar 
population and decreases as the cluster evolves and the stellar age increases.
No significant trend was found between these two variables
for the bulk of our sample, but a clear trend is visible for those
with the highest quality  EW(H$\beta$) and  O/H determinations. 
This result suggests that the presence of a range of ages in the data 
may be responsible for part of the scatter in the N2 vs.~O/H diagram.
This trend is shown in Figure 3 where we plotted the distance to the
regression line in Figure 1 versus the EW(H$\beta$).
There is a clear trend suggesting that part of the
scatter is correlated with EW(H$\beta$).
Intrinsic variations in
the  N/O abundance ratio will of course constitute another source of scatter.

\begin{figure}
\vspace{8.0 cm}
\includegraphics{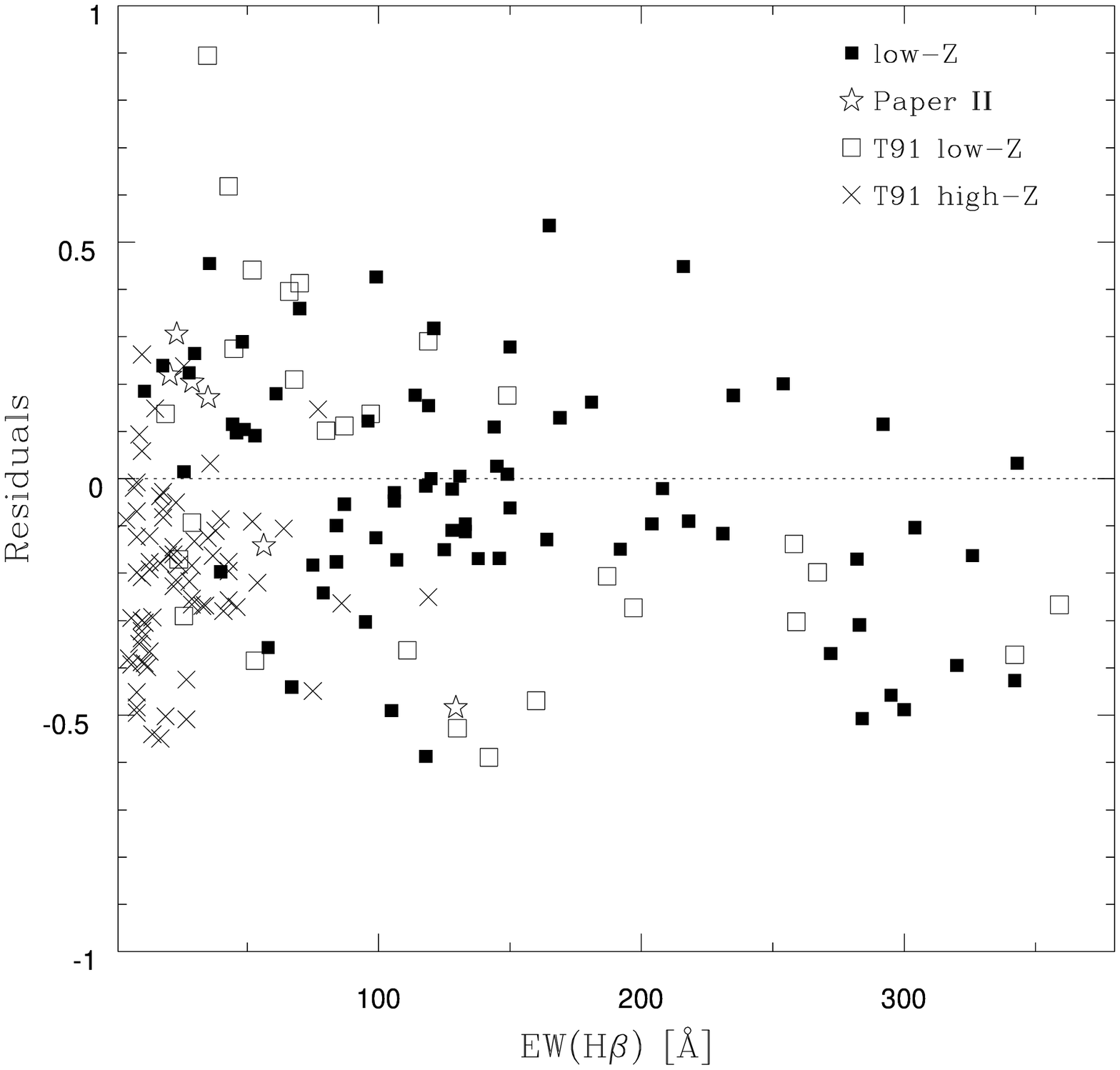}
\caption{Residuals of the linear regression shown in Figure 1 against
equivalent width of H$\beta$. The low metallicity sample can be found in Table 1.}
\end{figure}

We can also expect dispersions of 
the order of 0.05-0.2 dex in metallicity due to the differences 
between global spectra and smaller aperture exposures (Kobulnicky, 
Kennicutt \& Pizagno 1999). This difference  becomes more 
important when comparing low redshift 
galaxies with high-z spatially integrated spectra. 
Several effects my cause the emission-line ratios to produce oxygen abundance 
estimates significantly lower than those derived from small aperture 
observations of individual \Hii regions. Kobulnicky \etal (1999) demonstrated 
through simple modeling, that temperature fluctuations are the primary cause 
for overestimating the electron temperature and underestimating the oxygen 
abundance, and that ionization parameter variations further exacerbate this 
systematic underestimation. We therefore consider our emission line 
measurements and abundance determinations from global spectra as lower limits. 
A more careful analysis of spatial variation effects on emission line 
intensities will be carried out in Paper II.

The N2 parameter has clear observational advantages for ranking metallicities 
in starforming galaxies. Besides it being single-valued, it also appears to 
have a tighter correlation with O/H than  R$_{23}$; still, it has the drawback 
of using forbidden lines of  N, making the abundance calibrator 
sensitive to variations on the N/O abundance ratio. Photoionization models 
taking into account {\it primary plus secondary} nitrogen encompass the data 
in the N2 vs.~12+log(O/H) plane. To reproduce the dispersion 
it is enough to vary the ionization parameter {\it U} in the models by (+0.5,-1.0), with 
an average fit of {\it log(U)} $\cong$ -2.0. 

We can expect that for the very low metallicity galaxies, 
the nitrogen lines would become less sensitive to metallicity 
due to the increasing presence of primary N in comparison to secondary 
N. Although we do not clearly see this behavior in our sample, we find that 
an alternative line ratio for the low metallicity branch would be useful, 
as the \Nii lines also become too faint and difficult to deblend from 
H$\alpha$ once in this regime. We adopted the ratio 
log(\Sii$\lambda\lambda$6717,6731\AA /H$\alpha$) (S2) as an alternative.
In Figure 4 we show the relation of S2 with metallicity.  
This relation still has the advantage of not being 
strongly dependent on reddening corrections, 
but its scatter is apparently
larger than for the N2 -- 12+log(O/H) relation. 

We emphasize that the N2 relation is clearly not better than the
S$_{23(4)}$ methods, although it is observationally much easier to obtain,
specially for moderate redshift objects.
It therefore has the advantage of allowing a very fast ranking of galaxies in 
a metallicity sequence, in particular when searching for low metallicity 
objects.  A test for this method was performed on the Durham redshift survey 
data and
is the subject of Paper~II.

\begin{figure}
\vspace{8.0 cm}
\includegraphics{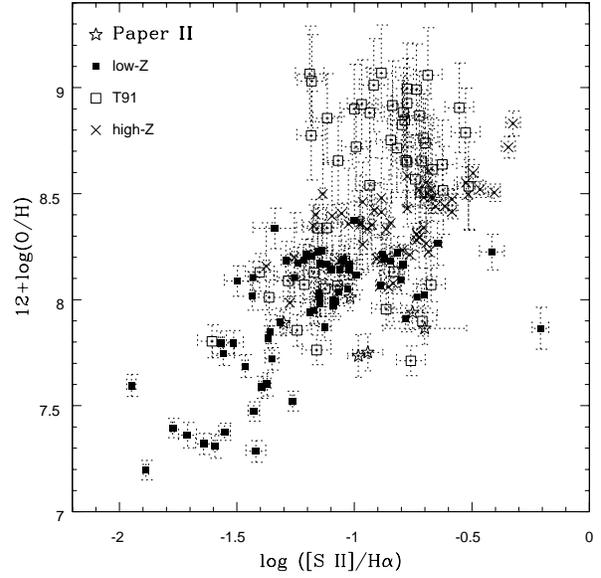}
\caption{Oxygen abundance versus intensity ratio of the 
emission lines  \Sii$\lambda$6717\AA \  +  \Sii$\lambda$6731\AA \   over 
H$\alpha$. The high metallicity sample is from the compilations of 
D\.az \& P\'erez-Montero (2000)
and Castellanos, D\.az \& Terlevich (2001); 
the low metallicity sample is found in Table 1.}
\end{figure}


\section{Conclusions}

We have collected from the literature
a representative sample of spectroscopic measurements of starforming galaxies
covering a wide range in metallicity  
(7.2 \aless 12+log(O/H) \aless 9.1), and recalculated oxygen 
abundances in a self-consistent manner. We confirmed previously published
results on the correlation between the logarithmic ratio of 
\Nii/H$\alpha$ (N2) and O/H and obtained an improved 
calibrator thanks to a larger sample of carefully and 
consistently calculated abundances and line ratios.

Considering the weak points and limitations of empirical
abundance determination methods, we reckon that  using N2 as 
a metallicity calibrator presents several advantages: 
it involves easily measurable lines
that are available for a large redshift range (up to $z \sim$2.5);  the N2 
vs.~metallicity relation
is monotonic;  the \Nii and H$\alpha$ lines
can be separated in 
even moderate resolution spectra, and  the N2 line ratio does not
depend on reddening corrections or flux calibration.

On the negative side, N2 is sensitive to ionization and O/N 
variations implying that, strictly, it should be used mainly
as an indicator of galaxy-wide abundances.

Interestingly, the comparison with photoionization models 
indicates that the observed N2
is consistent with  nitrogen being a combination of both primary and 
secondary origin.

Our main conclusion is that the combination of N2 and S2 provides a tool
to roughly map the metallicity of galaxies (and search for low Z galaxies)
using survey quality data, like e.g.~that of the Sloan Digital Survey, for
redshifts in excess of 0.2.

\section*{Acknowledgments}

We are grateful to Marcelo Castellanos and Angeles D\.az for providing us
data prior to publication.
GD would like to thank CNPq-Brazil for a research scholarship, and INAOE for
hospitality during a visit to Mexico where part of this paper was written.
ET acknowledges support through a research grant from CONACYT-Mexico
and everlasting hospitality of IoA.




\appendix


\label{lastpage}
\end{document}